# FlowMLLab: An open-source framework for scientific machine learning in thermal-fluid and aerospace engineering

Ehsan Roohi

*Department of Mechanical and Industrial Engineering, University of Massachusetts Amherst, Amherst, Massachusetts, United States*

ABSTRACT

FlowMLLab is an open-source Python framework for scientific machine learning in thermal-fluid and aerospace engineering, covering incompressible, compressible and rarefied flows. Its 22 weeks comprise 51 notebooks and 35 lectures. Solvers and molecular sampling underpin field prediction, uncertainty quantification and reduced-order modeling. Proper orthogonal decomposition (POD) and deep operator networks (DeepONets), including POD-DeepONet, connect operating conditions to flow fields alongside Fourier neural operators. Physics-informed neural networks infer solutions and constitutive coefficients. Geometry-conditioned models, including the point-cloud DeepONet variant Geom-DeepONet, handle changing domains. Aerospace applications address airfoil shock/vortex identification, rarefied-gas modeling, nozzle flow prediction and supersonic body design. Complementary thermal-fluid examples cover unsteady wakes, hydrofoil vapor clouds and noisy particle observations. Attention mechanisms support sparse-sensor reconstruction, autoregressive prediction and transfer assessment, alongside character-level language modeling. Complete-case evaluation excludes tests from selection; execution notes identify retained research outputs. A worked extension embeds a learned monotone relation for a fixed manufactured Carreau fluid in channel momentum balance. At three unseen in-support pressure gradients, it outperforms profile-interpolation and fitted-Newtonian controls with different information or priors.

## Code metadata

Table 1 summarizes software metadata.

**Table 1**
Software metadata and reproducible experiment sources.

| Item | Record |
|---|---|
| C1 Version | FlowMLLab v1.10.0. |
| C2 Permanent source | Immutable source `68a89e1`. |
| Digital object identifier (DOI) | Tag `89cb54c`; 10.5281/zenodo.23210097. |
| C3 License; C4 versioning | Massachusetts Institute of Technology (MIT) license; Git; separate data permissions. |
| C5 Languages/libraries | Python, Jupyter, NumPy, SciPy, pandas, Matplotlib, scikit-learn; optional TensorFlow/Keras and PyTorch backends. |
| C6 Environment | Python ≥ 3.10, < 3.14; extras in `pyproject.toml`; Linux, macOS or Windows. |
| C7 Documentation | Setup, course map and provenance. |
| C8 Contact | roohie@umass.edu. |

## 1. Motivation and significance

Scientific machine learning connects computational fluid dynamics (CFD) with artificial intelligence (AI), including physics-informed neural networks (PINNs) [1,2]. In teaching, implementing a network is only part of the task. Students must establish what the numerical reference

*Email address:* roohie@umass.edu.

resolves, decide which cases may inform training, and check whether a small field error preserves the relevant flow behavior.

FlowMLLab addresses these tasks through a connected set of thermal-fluid and aerospace experiments. Its 22 weeks and companion laboratories progress from numerical methods and statistical learning to kinetic surrogates, machine vision, geometry-dependent operators and attention. Developed for the graduate course *AI in Fluid Mechanics* at the University of Massachusetts Amherst, the collection supports a six-week core and selectable extensions. Lectures introduce assumptions; notebooks expose the corresponding calculations. The contribution is a reusable workflow joining physical references, model construction and diagnostics, demonstrated by the constitutive channel extension in Section 4.2.

## 2. Software and teaching structure

### 2.1. Architecture and execution

`flowmllab` and `common` separate physical cases from learning. Records retain coordinates, fields, boundary conditions and parameters. Adapters construct inputs or reduced representations; diagnostics evaluate the declared physical support.

The course website provides prerequisites, installation routes and notebook guides. Table 2 links to each weekly guide; the pinned inventory records 51 notebooks and 35 lectures [3]. Three execution routes distinguish between inspecting retained evidence, running a compact teaching calculation and regenerating a complete protocol. Units specify central processing unit (CPU) or graphics processing unit (GPU) requirements and research-data/solver dependencies.

### 2.2. Data and comparison contracts

Case contracts fix the physical support and complete-case splits; tests are excluded from preprocessing and selection. Cavity scales are $L$, $U_{lid}$ and $\rho U_{lid}^2$; $\rho$ denotes density and Re the Reynolds number. Pressure has a declared gauge.

The field error uses the relative Euclidean norm

$$\epsilon_2(q) = \frac{\|q_{pred} - q_{ref}\|_2}{\|q_{ref}\|_2}. \tag{1}$$

For spatial maps, root-mean-square (RMS) reference scaling gives

$$e_q(\mathbf{x}) = 100\frac{q_{pred}(\mathbf{x}) - q_{ref}(\mathbf{x})}{q_{\text{RMS}}}, \qquad q_{\text{RMS}} = \sqrt{\frac{1}{N}\sum_{i\in\Omega} q_{ref,i}^2}. \tag{2}$$

Here $\Omega$ is the stated comparison support and $N$ its sample count. One reference scale avoids division by local zeros; RMS($e_q$) equals $100\epsilon_2$. Maps retain signs unless marked absolute, common comparison scales and all extrema. Zero reference norms are undefined. Profiles, boundary checks and constitutive admissibility supplement field errors.

## 3. Methods across the collection

Table 2 introduces the full collection before the methods are developed by topic.

### 3.1. Numerical, statistical and reduced-order foundations: weeks 1–6

Week 1 develops arrays and automatic differentiation, then develops a finite-difference vorticity–streamfunction solver, before comparing velocity and pressure with independent references [4–6]. Finite-volume (FV) pressure coupling covers the semi-implicit method for pressure-linked equations (SIMPLE), pressure-implicit splitting of operators (PISO) and the PISO–SIMPLE hybrid (PIMPLE). Week 2 constructs features, complete-case splits and regression baselines: k-nearest neighbors (KNN), deep neural networks (DNNs) and Bayesian prediction. Proper orthogonal decomposition (POD) [7] with Gaussian-process (GP) regression introduces uncertainty calibration. Week 3 samples Maxwellian velocities, recovers molecular moments and implements direct simulation Monte Carlo (DSMC) transport and collisions [8].

Week 4 generates qualified field data and compares multilayer perceptrons (MLPs) with a POD-based deep operator network (POD-DeepONet): a scalar Reynolds-number branch and a fixed POD trunk [9,10]. POD–Galerkin and the discrete empirical interpolation method (DEIM) approximate dynamics [11]. A Fourier neural operator (FNO) learns Stokes-to-Navier–Stokes corrections (Fig. 1) [12]. Weeks 5–6 test physical penalties, sensing, representation rank, data sufficiency and kinetic closures with frozen protocols.

**Table 2**
Course map: 22 weeks and 10 numbered companions. Each unit links to its lecture/notebook guide at `68a89e1`; the course website provides the catalog [3].

| Unit | Methods and learning task |
|---|---|
| 1 | Python, automatic differentiation and independently validated cavity-flow calculations. |
| 1.1 | Scientific-code generation checked through manufactured solutions and numerical audits. |
| 1.2 | Finite-volume pressure–velocity coupling and independent cavity pressure reconstruction. |
| 2 | Rarefaction regression with complete-case splits and supervised learning baselines. |
| 2.1 | Bayesian prediction and uncertainty calibration for reduced flow representations. |
| 3 | Maxwellian sampling, molecular moments, particle transport and binary collisions. |
| 4 | Qualified CFD datasets for scalar, field and operator-learning surrogates. |
| 4.1 | Projection-based cavity dynamics, nonlinear hyper-reduction and computational cost assessment. |
| 4.2 | Physical Stokes-to-Navier–Stokes correction across Reynolds numbers and lid profiles. |
| 5 | Controlled projects on physical penalties, sensing and data sufficiency. |
| 6 | Project validation through stochastic sampling, rarefaction transfer and kinetic closures. |
| 7 | Cylinder-wake simulation, convolutional field prediction and modal time forecasting. |
| 7.1 | Data-based surrogates and interpolation for rarefied hypersonic cylinder fields. |
| 7.2 | Noisy wake assimilation using reduced dynamics and Kalman filtering. |
| 7.3 | Masked wake pretraining with matched reconstruction and label-efficiency controls. |
| 7.4 | Diverse-wake representation transfer to lift prediction under target-label budgets. |
| 8 | Exact compressible-flow relations and branch-aware neural inverse inference. |
| 9 | Zonal micro-step surrogates, shock-aligned nozzles and data-alignment audits. |
| 10 | Data-driven rarefied cavity and molecular-shock surrogates with retained references. |
| 10.1 | Classical scattering calculations, collision-angle surrogate audits and research companions. |
| 11 | Shock/vortex identification, velocity reconstruction and hydrofoil vapor-cloud detection. |
| 12 | Noise-aware reconstruction of molecular moments from independently sampled observations. |
| 13 | Hard-constrained cavity networks with residual checks and regional flow assessment. |
| 14 | Inverse diffusion inference and neural coefficients for coupled turbulence closures. |
| 15 | Geometry-conditioned flow operators evaluated on excluded double-step channel families. |
| 16 | Surrogate-guided supersonic shape design with pressure and drag checks. |
| 17 | Constructing query/key/value attention and interpreting its learned-kernel computations. |
| 18 | Character-level language modeling with likelihood and causal-prefix verification. |
| 19 | Flow tokenization, sensor design and reduced-state representation limits. |
| 20 | Attention-based reconstruction with missing sensors, noise and classical controls. |
| 21 | Autonomous flow rollout evaluated against modal and neural forecasting baselines. |
| 22 | Transfer assessment accounting for pretraining, adaptation labels and optimization budgets. |

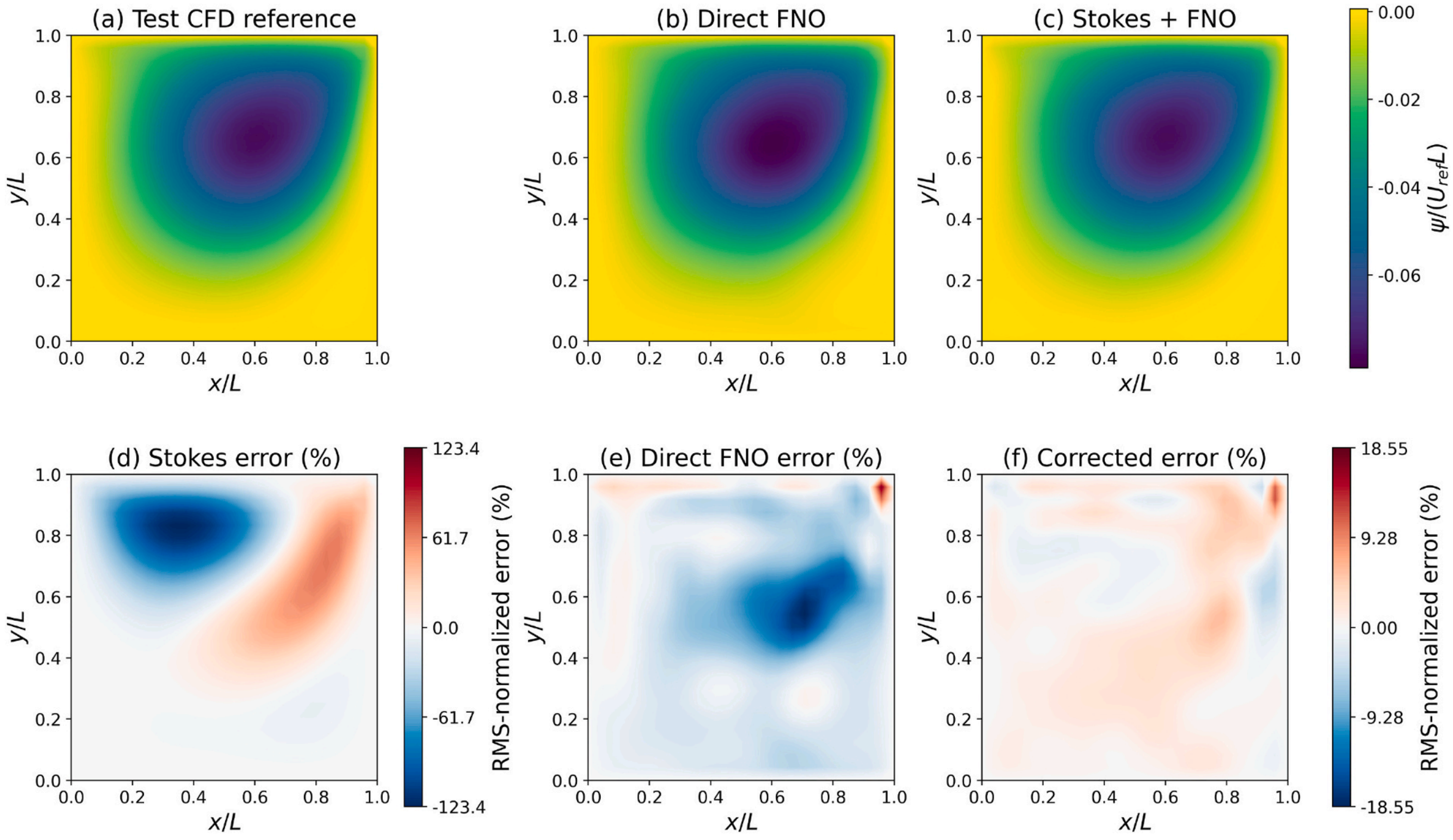

**Fig. 1.** Multifidelity reconstruction. Training/validation/test: 16/4/6 diverse-lid cases (training Re = 144.5–376.8); shown: the retained test Re = 378.4, seed 17. Top: CFD, direct FNO and Stokes-corrected FNO streamfunction; bottom: signed percentage errors, Eq. (2). Contours interpolate from unchanged $25 \times 25$ nodes. Top fields and FNO errors share their respective scales. $L, U_{ref}$ denote length and speed; $\mathrm{Re} = U_{ref} L/\nu$ [3].

### 3.2. Dynamics and compressible/kinetic models: weeks 7–10

The lattice Boltzmann method (LBM), with a two-dimensional nine-velocity (D2Q9) lattice, supplies cylinder wakes. Convolutional neural networks (CNNs) predict subsequent fields (Fig. 2). Sensing compares POD–Kalman with persistence and dynamic mode decomposition (DMD) [13]; masked-autoencoder pretraining tests adaptation against gappy-POD controls.

Exact gas-dynamic relations expose branch ambiguity before inverse learning [14]. Micro-step fits use zonal MLPs alongside DeepONet construction exercises motivated by the rarefied step study [15]. Nozzle examples compare physical and shock-aligned representations [16]. POD–polynomial shock models (Fig. 3) use attributed DSMC data [17, 18]; neural collision-map studies supply separately identified research evidence [19].

### 3.3. Vision and noisy observations: weeks 11–12

Manufactured controls introduce shock/vortex identification. A U-shaped encoder–decoder (U-Net) reconstructs velocity before swirling-strength calculation; a direct mask predictor bypasses reconstruction [20]. A frozen shared-encoder model identifies shocks and vortex cores on retained airfoil fields (Fig. 4) [21,22]. Hydrofoil companions examine vapor-cloud detection.

Noise2Noise [23] patch-MLPs denoise DSMC moments. Geometry-native reconstruction with support monitoring supplies related research context [24].

### 3.4. Physics constraints, geometry and design: weeks 13–16

The streamfunction PINN uses automatic differentiation and hard-wall constraints [25]. Derivative and residual checks accompany training. A retained checkpoint and Nektar++ CFD provide streamlines, streamfunction, vorticity and absolute dimensionless differences (Fig. 5). Regional assessment of a depth/width-five cavity further illustrates why streamline geometry and rotational-field discrepancies should be examined together [26]. An inverse PINN fits a manufactured closure, while Reynolds-averaged Navier–Stokes (RANS) companions audit coupled-flow evidence [27].

Geometry studies use DeepONet, its point-cloud Geom-DeepONet variant [28] and geometry-aware Fourier neural operator (Geo-FNO) [29]. Alternatives include Scalable Mesh-free Aerodynamic Simulations from Raw Geometries using a Transformer-based Surrogate Model (SMART) [30], geometry-aware physics-attention transformer GeoTransolver [31] and decomposable multiscale iterative neural operator (DoMINO) [32]. Authors' code uses case adapters, planar configurations and validation-only tuning. References are sampled OpenFOAM CFD fields; the producing solver version and complete case configurations are not recovered. Fig. 6 tests a double-step family excluded from training and validation. Training geometries comprise recesses, internal obstacles, single steps and smooth/ramped expansions.

Design laboratories use ridge, MLP and GP surrogates to minimize peak $C_p$ at $r/L = 0.5$ for a fixed-volume axisymmetric body, allowing at most 2% greater pressure drag than the baseline. The body has length $L = 1$ m and volume $\pi(0.06)^2/2$ m$^3$. Gmsh meshes its meridional plane; Stanford University Unstructured (SU2) 8.0.1 solves axisymmetric compressible Euler flow. Fig. 7 compares independently computed pressures for different bodies: the optimized red curve lowers the compression peak while deepening the aft expansion. It does not measure atmospheric propagation or ground-level loudness.

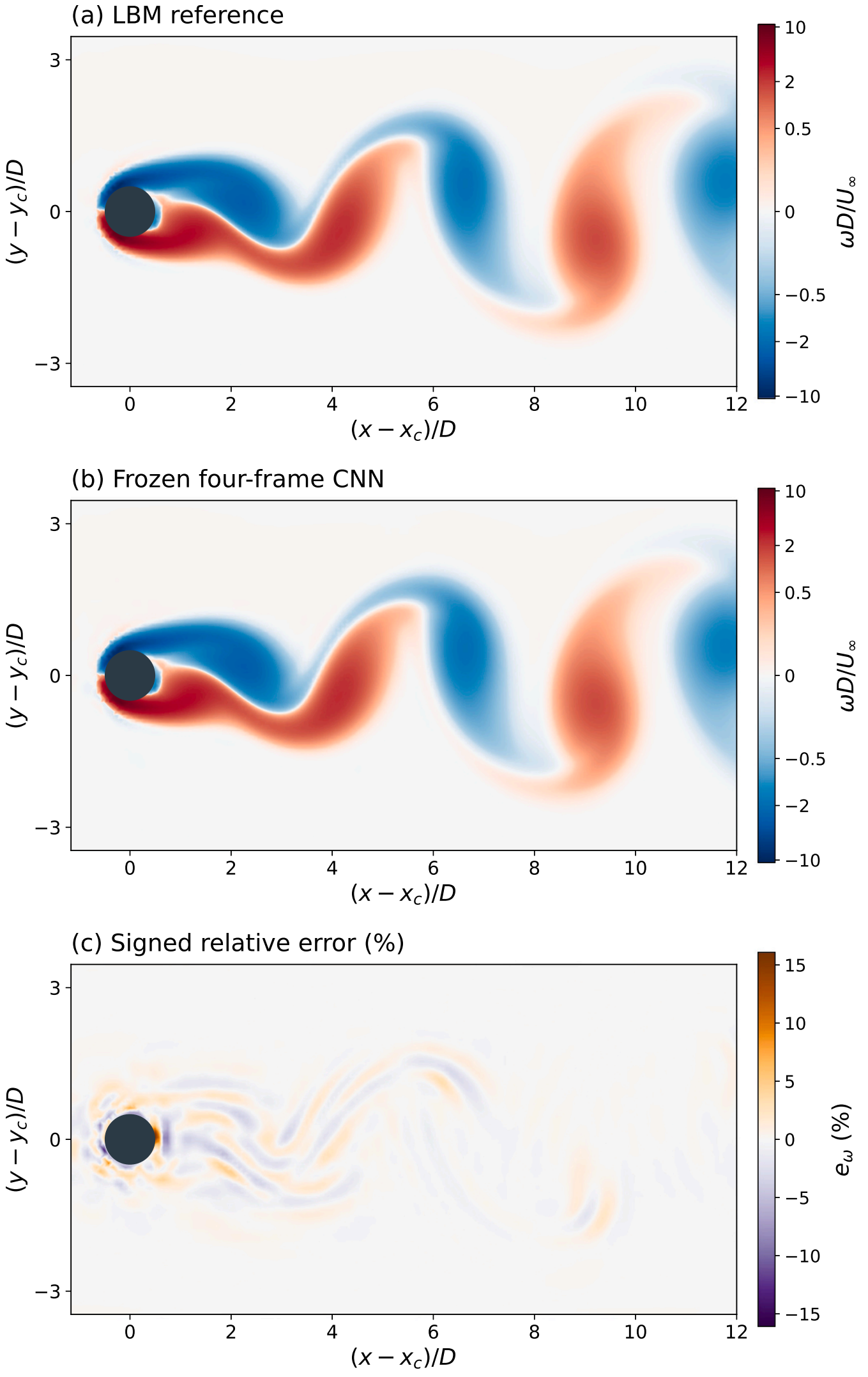

**Fig. 2.** One-step wake prediction. CNN training: Re = 60, 80, 90, 110, 120, 140; validation: 100; test: 105. Four observed test frames 138–141 are used to predict frame 142 ($tU_\infty/D = 77.29$). Panels show LBM, frozen CNN and signed percentage vorticity error, Eq. (2). Native-node contours retain extrema; vorticity uses a symmetric-log scale and the error a linear scale; $D, (x_c, y_c)$ denote the cylinder diameter and center [3].

### 3.5. *Attention and sequence representations: weeks 17–22*

Students construct query/key/value attention [33] before training a small character-level Transformer. Prefix checks establish causal decoding. Flow exercises examine tokenization and POD representation floors.

Sensor-set attention reconstructs fields from sparse observations in a declared wake region of interest (ROI), with sensor removal and gappy-POD controls (Fig. 8). Rollout uses MLP/DMD controls. Transfer accounts for pretraining, labels and optimization budgets. These teach language-model mechanisms, not a fluid foundation model.

## 4. Numerical evidence and executable reuse

### 4.1. *Evidence and execution scope*

Manifests distinguish fitting from retained research. Figure provenance records source hashes, case identities and deterministic plotting operations. The Fokker–Planck companion uses attributed data [34], without reproducing that study's closed-loop calculation.

### 4.2. *Worked constitutive extension*

The reuse example solves the steady flow of a fixed Carreau material between plates at $y = \pm 1$ [35]. With $G = -\mathrm{d}p/\mathrm{d}x > 0$, momentum gives $\tau_{xy} = -Gy$ and

$$\mu(\gamma) = \mu_\infty + (\mu_0 - \mu_\infty)[1 + (\lambda\gamma)^2]^{(n-1)/2}, \quad |\tau| = \mu(\gamma)\gamma, \tag{3}$$

where $\mu_0 = 1$, $\mu_\infty = 0.1$, $\lambda = 2$, $n = 0.5$. All quantities are nondimensional. Bracketed roots and independent quadrature supply references. A monotone piecewise cubic Hermite interpolating polynomial (PCHIP) learns $\gamma = f(|\tau|)$. Extrapolation is rejected; odd reflection preserves signed shear and dissipation. Deployment gives

$$u(y) = \int_{|y|}^{1} f(Gs)\,\mathrm{d}s, \qquad Q = 2\int_{0}^{1} s f(Gs)\,\mathrm{d}s. \tag{4}$$

Training uses $G = 0.5, 1.5, 3$, validation uses $G = 1$, and evaluation $G = 0.75, 2, 2.75$. A predeclared validation-only 0.1% velocity/flux rule selects five nodes/case (ten unique labels) from 5/9/17/33, without evaluation retuning. Newtonian fitting uses those labels; interpolation uses three complete profiles. Their information budgets and priors differ.

Fig. 9 and Table 3 retain every evaluation case. Walls are exact; asymmetry is below $2.3 \times 10^{-15}$. Stresses remain within training support; material extrapolation is untested.

The adapter reuses error and fingerprint utilities.

### 4.3. *Verification and limits*

Five extensions and twenty selected core tests pass, including Newtonian-limit, refinement, admissibility, wall/symmetry and split checks; repeated arrays match.

Sampling, retrospective geometry selection and limited vision references bound interpretation.

## 5. Reuse, impact and conclusions

Aerospace applications distinguish aerodynamic shock/vortex identification and supersonic body design (Figs. 4 and 7) from gas-kinetic shocks, collisions and nozzle surrogates for low-density flight and internal propulsion [16,17,19]. Thermal-fluid examples address channels, cavities and hydrofoils. Cases share data/evaluation interfaces while retaining their own governing equations and reference solvers.

Shared case records and frozen protocols support graduate teaching and research. The channel extension demonstrates reuse under changed geometry and rheology. Cited gas-dynamics and collision studies provide research provenance [14,18,19]; documented course use supports educational relevance. Measured learning gains and independent published adoption have not been established.

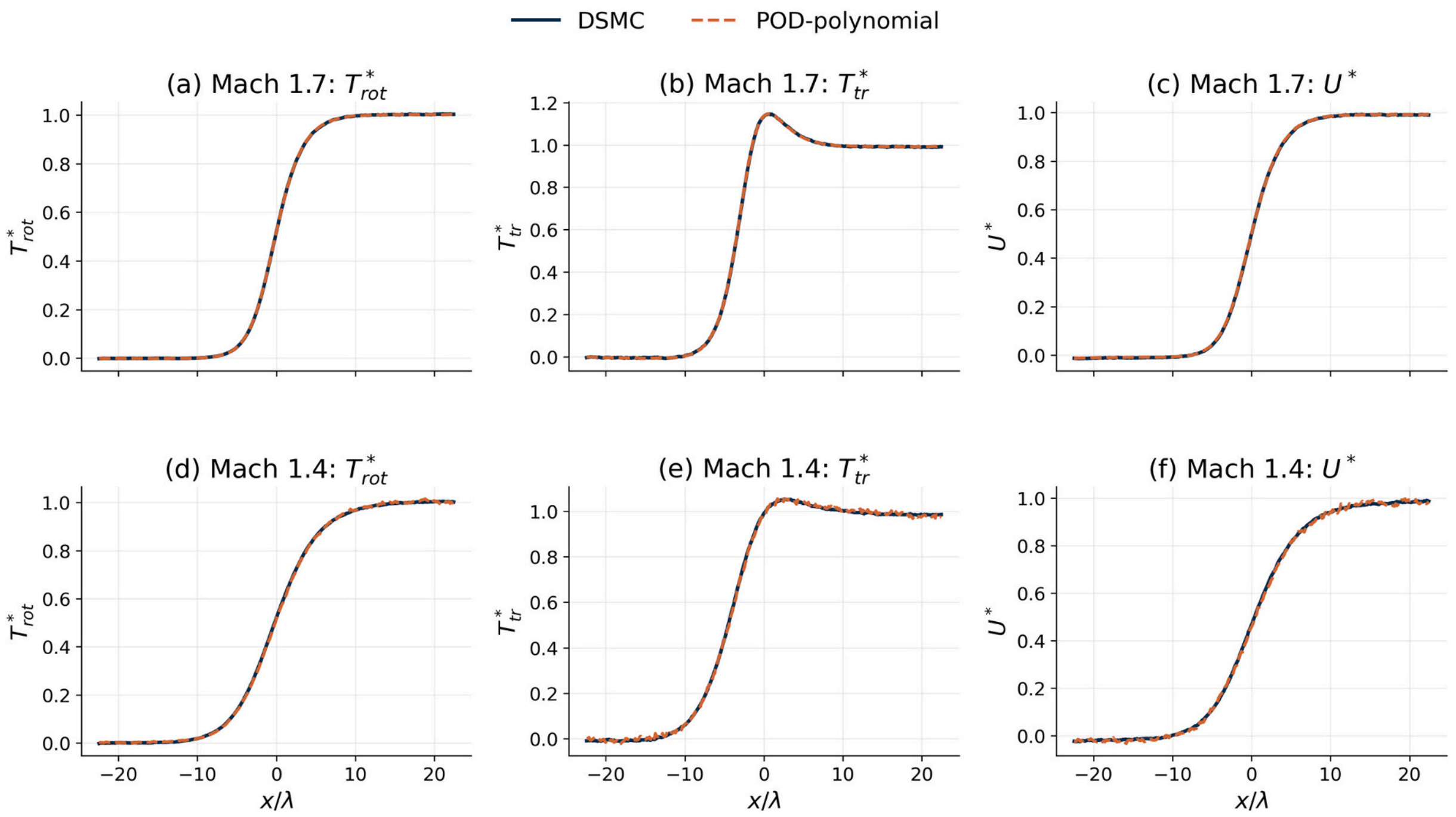

**Fig. 3.** Diatomic-shock DSMC and POD-polynomial predictions, fixed rank 4/degree 3. Top: training Mach 1.4/1.5/1.6/1.8/1.9, test 1.7; bottom: training 1.5–1.9, test 1.4; no validation tuning. Starred quantities retain source normalization; $\lambda$ is the mean free path [3,17].

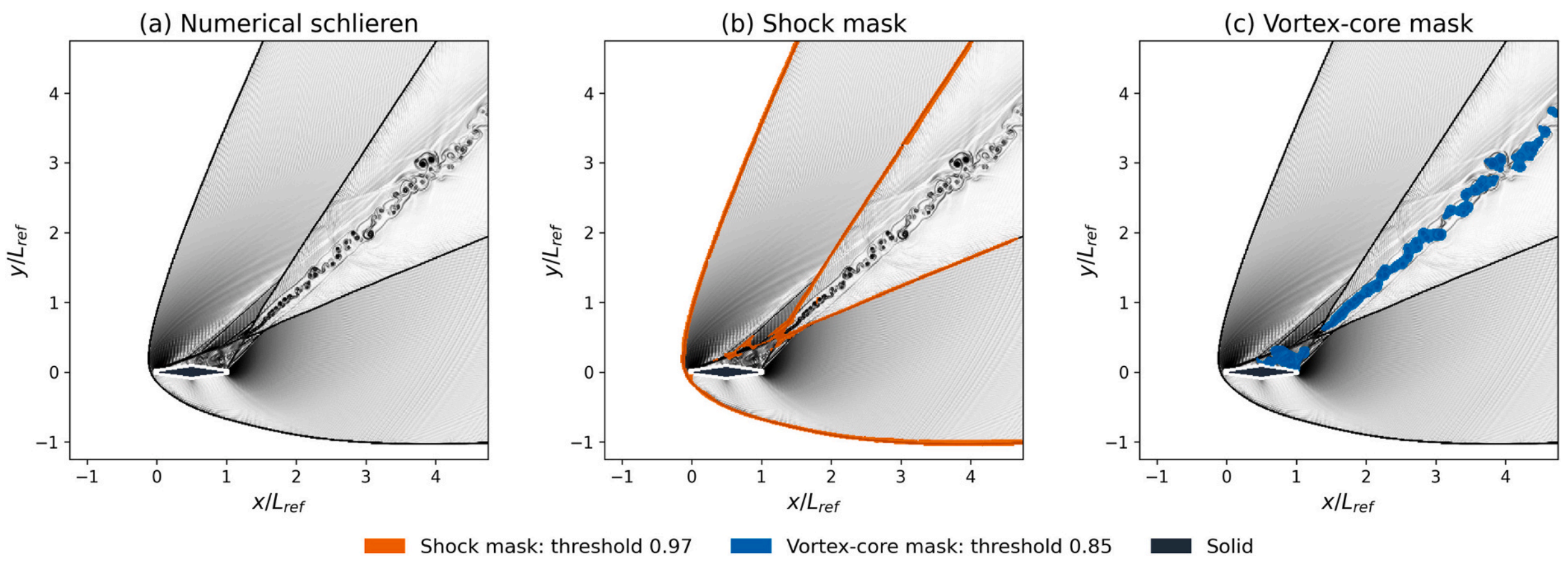

**Fig. 4.** Diamond-airfoil development visualization at $t = 28.35$: density schlieren, shock mask and vortex-core mask from the frozen research checkpoint. Upstream training/calibration frame IDs are unarchived; this is not an independent test. Legend thresholds of 0.97/0.85 are sigmoid scores, not accuracies. $L_{ref}$ is the retained length scale [21,22].

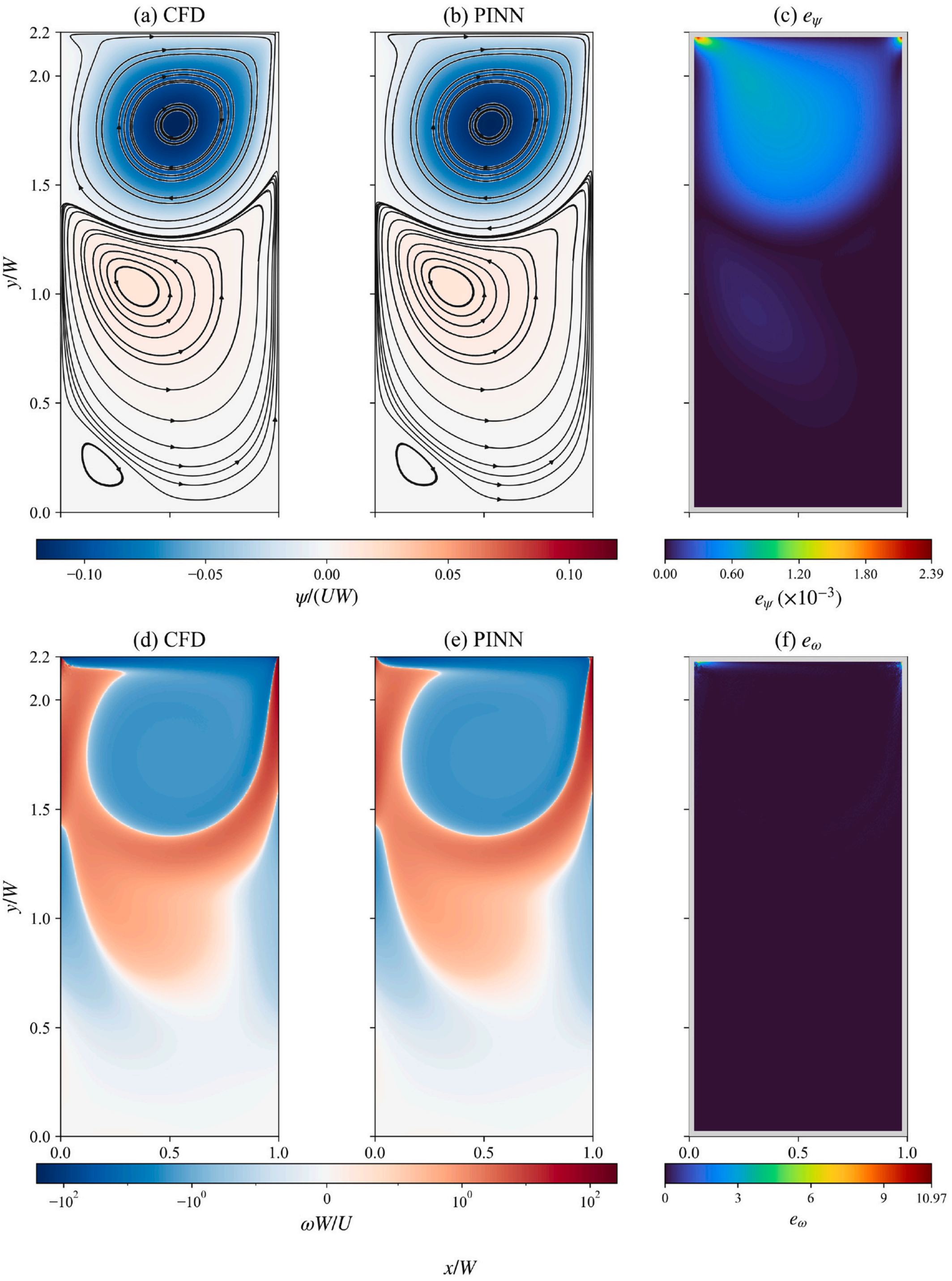

**Fig. 5.** Physics-trained PINN versus independent Nektar++ CFD, Re = 1000, $D/W = 2.2$. Top: streamfunction and velocity-integrated streamlines; bottom: sampled vorticity. Absolute dimensionless errors are $e_\psi = |\psi_{\text{PINN}} - \psi_{\text{CFD}}|/(UW)$ and $e_\omega = |\omega_{\text{PINN}} - \omega_{\text{CFD}}|W/U$ [26], Eq. (29), with analogous streamfunction scaling, not percentages. Gray excludes the 0.02 W wall band; linear error scales are unclipped. Field scales/seeds are shared; vorticity uses symmetric-log scaling. $U, W$ denote the lid speed and width. Post-training comparison; collocation coordinates unarchived; retained lid profiles differ.

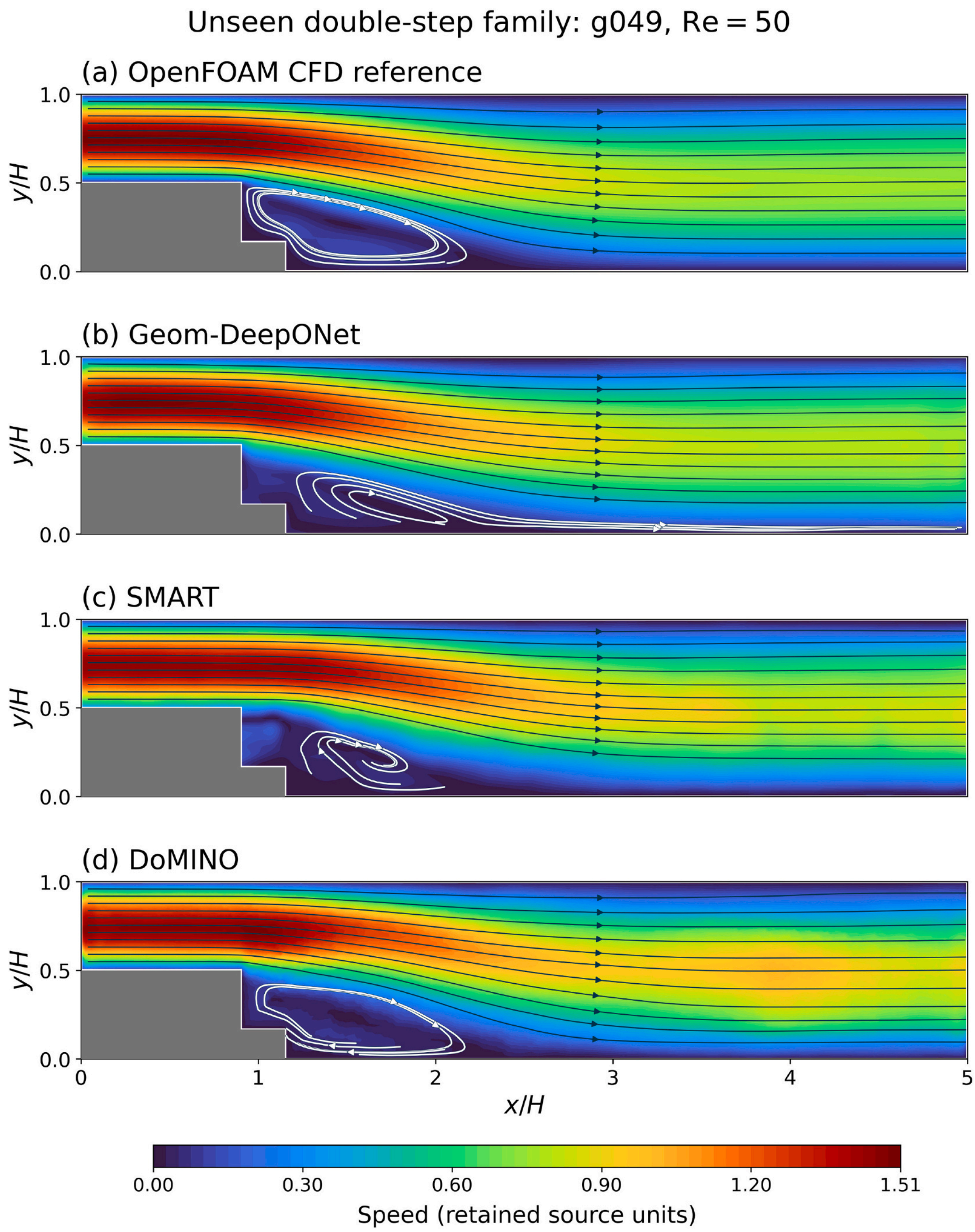

**Fig. 6.** Unseen double-step g049, Re = 50: CFD and archived neural predictions. Training: 100 cases on 39 geometries g001–g044 excluding g005/g007/g012/g018/g032; validation: 8; retrospective tests: 19. **The double-step family is absent from training and validation.** All panels use the same speed scale; trajectories use each panel's velocity and common seeds, without forced closure. Domain aspect ratio is 5 : 1; $H$ is channel height. Training masks appear in the separate geometry sheet [3].

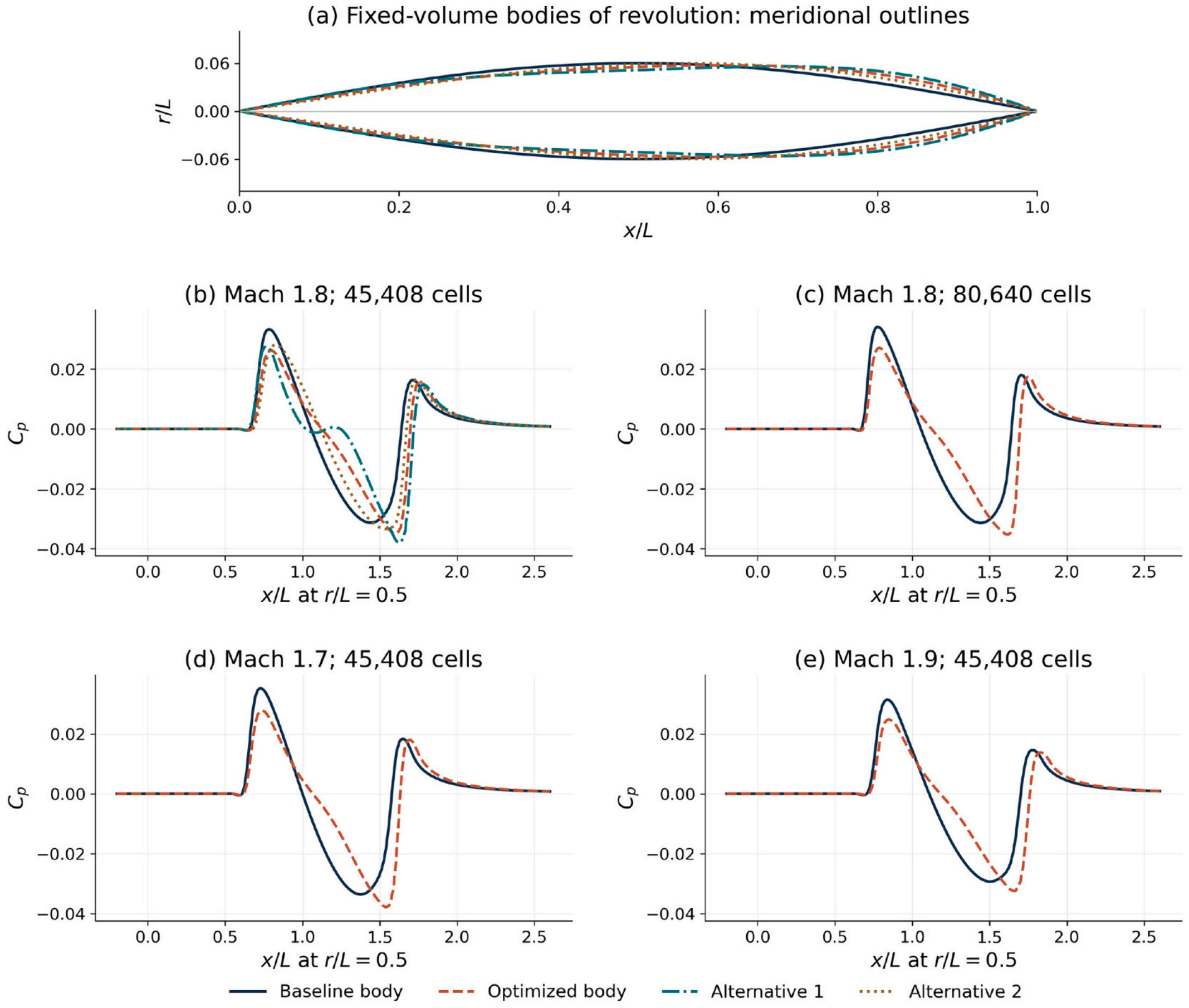

**Fig. 7.** Fixed-volume axisymmetric bodies: (a) meridional outlines; (b–e) direct SU2 CFD pressure profiles at $r/L = 0.5$. Historical selecting-surrogate training/validation/test/extrapolation used 24/6/8/6 geometries. Solid baseline and dashed red optimized curves represent different bodies, so they need not coincide; these are CFD reassessments, not prediction/reference pairs. $C_p = (p - p_\infty)/(\rho_\infty U_\infty^2/2)$; $L$ is body length [3].

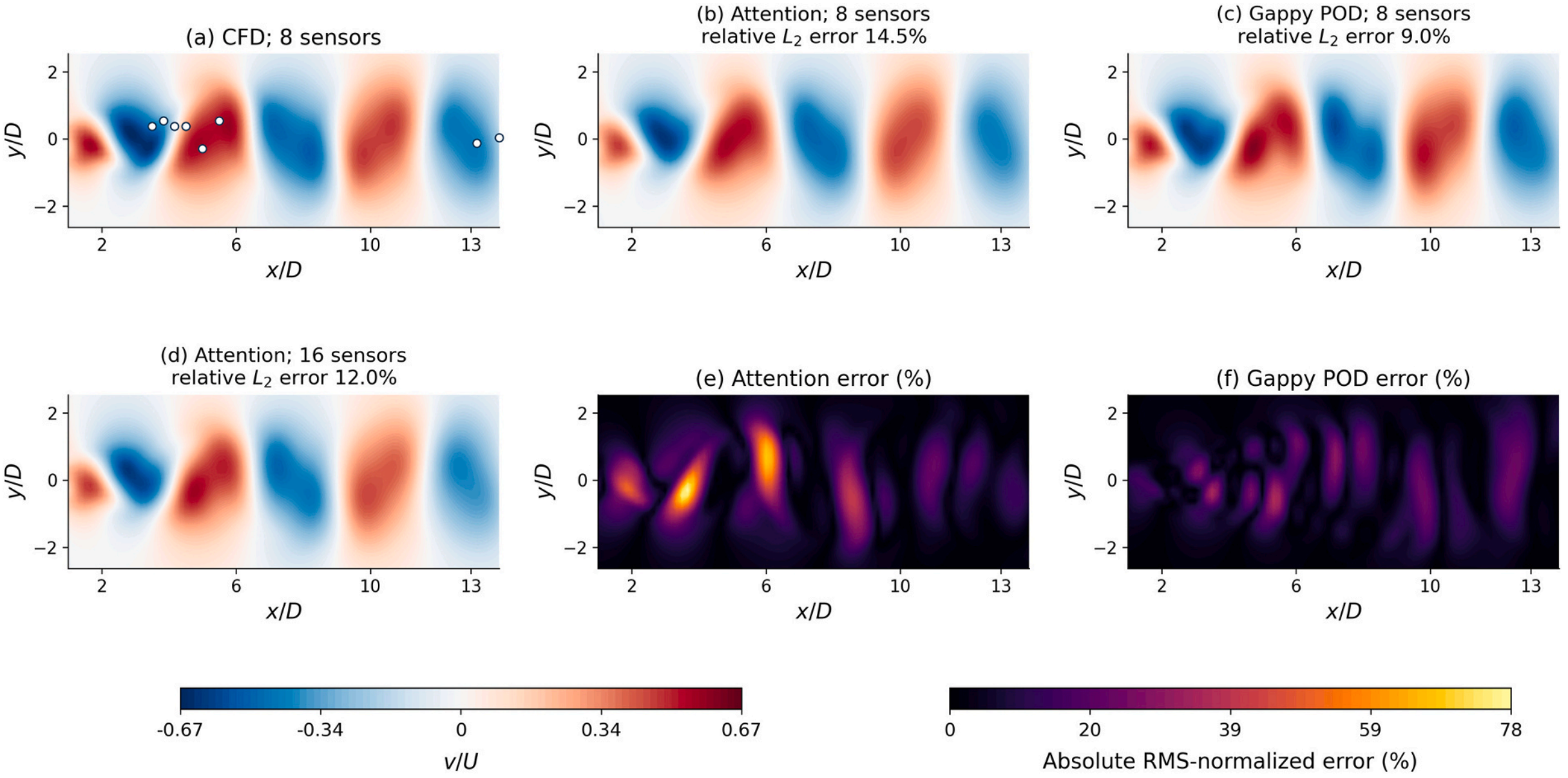

**Fig. 8.** Sensor reconstruction. Training: Re = 90, 110, 282 alternate-frame snapshots; validation: 100; test shown: 105, frame 140. CFD, attention and gappy POD use the same 8 of 16 observations in (b,c,e,f); (d) uses all 16. Panels (e,f): absolute percentage errors, Eq. (2), sharing a scale over the wake ROI [3].

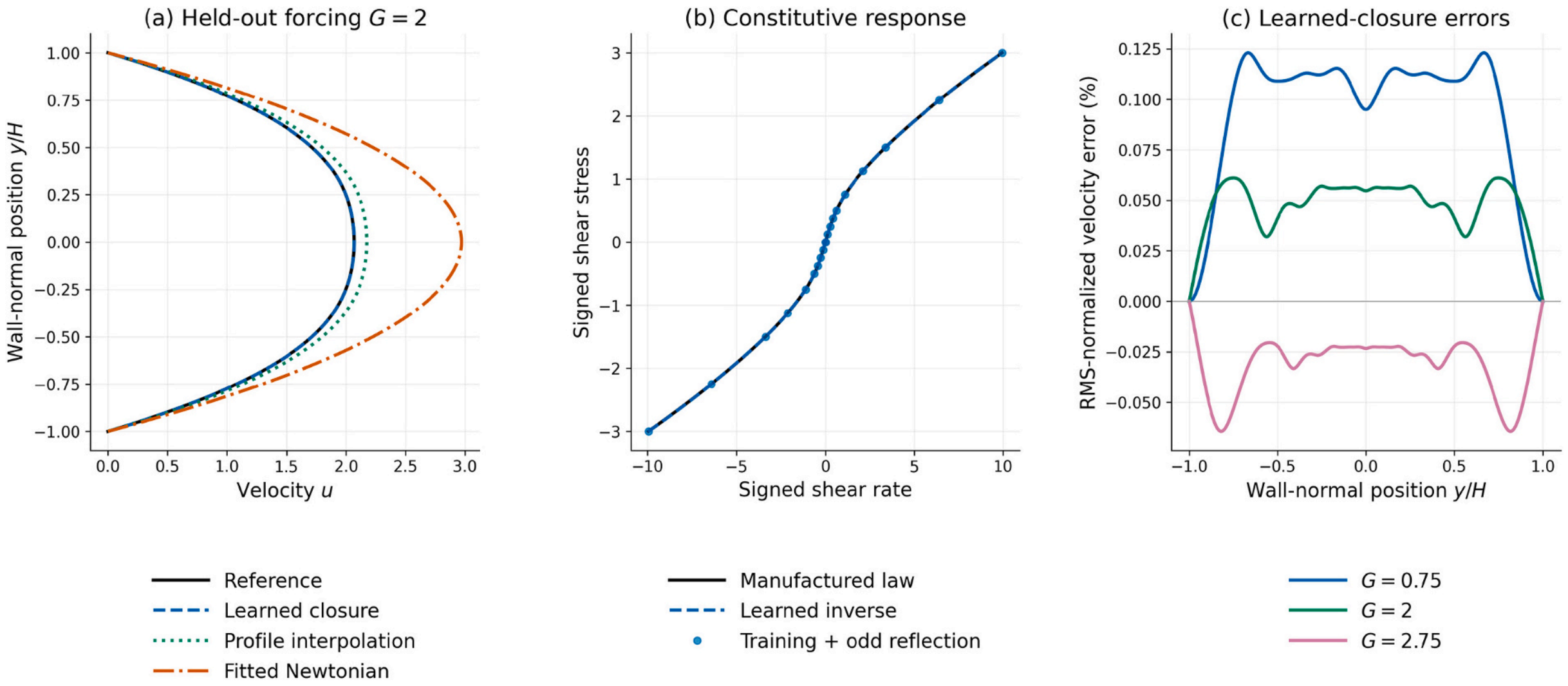

**Fig. 9.** Channel reuse. Training: $G = 0.5, 1.5, 3$; validation: 1; tests: 0.75/2/2.75. Panels: (a) held-out $G = 2$ velocity; (b) constitutive fit; (c) all held-out signed percentage errors, Eq. (2). Closure/Newtonian models use ten constitutive labels; interpolation uses complete training profiles. Walls: $y/H = \pm 1$, $H = 1$ [3].

**Table 3**
All held-out cases: velocity ($u$) and flux ($Q$) errors in percent. Profile interpolation uses complete training profiles; the Newtonian control fits the closure's ten labels.

| | Closure | | Profile interpolation | | Newtonian | |
|---|---|---|---|---|---|---|
| $G$ | $u$ | $Q$ | $u$ | $Q$ | $u$ | $Q$ |
| 0.75 | 0.10049 | 0.10240 | 15.94312 | 16.15810 | 131.61831 | 129.22704 |
| 2 | 0.04949 | 0.05172 | 5.16382 | 5.09902 | 36.18286 | 33.41084 |
| 2.75 | 0.03430 | 0.03385 | 1.75640 | 1.73303 | 14.10282 | 11.05882 |

## Declaration of Generative AI and AI-assisted technologies in the writing process

OpenAI ChatGPT/Codex assisted with code, deterministic plots and writing; Anthropic Claude Fable 5.1 and Opus 5.5 reviewed wording and figures. Fields derive from documented runs. The author reviewed and edited the output and takes full responsibility.

## Declaration of competing interest

The authors declare that they have no known competing financial interests or personal relationships that could have appeared to influence the work reported in this paper.

## Data availability

Sources: 68a89e1; release v1.10.0: 10.5281/zenodo.23210097. Separate data permissions apply.